\documentclass[a4paper]{jpconf}
\usepackage{graphicx}
\usepackage{pb-diagram,lamsarrow,pb-lams,amsfonts,amssymb,amsthm}
\usepackage{amsmath,epsfig}
\usepackage{cite}
\def\be{\begin{equation}}
\def\ee{\end{equation}}
\def\bea{\begin{aligned}}
\def\eea{\end{aligned}}
\def\ba{\begin{eqnarray}}
\def\ea{\end{eqnarray}}

\def\yzero{\smash{\hbox{$y\kern-4pt\raise1pt\hbox{${}^\circ$}$}}}

\def\-{\hphantom{-}}

\def\s2{\frac{1}{\sqrt2}}

\def\IF{\relax{\rm I\kern-.18em F}}
\def\II{\relax{\rm I\kern-.18em I}}
\def\IP{\relax{\rm I\kern-.18em P}}
\def\IC{\relax\hbox{\kern.25em$\inbar\kern-.3em{\rm C}$}}
\def\IR{\relax{\rm I\kern-.18em R}}

\def\Dsl{\,\raise.15ex\hbox{/}\mkern-13.5mu D} 
\def\IZ{Z\kern-.4em  Z}

\begin{document}
\title{On global aspects of duality invariant theories: M2-brane vs DFT}

\author{G. Abell\'an$^{2,a}$, C. Las Heras$^{1,b}$, M.P. Garc\'{\i}a del Moral$^{1,c}$, J.M. Pe\~na$^{2,d}$, A. Restuccia$^{1,3,e}$}

\address{$^1$Departamento de F\'{\i}sica, Universidad de Antofagasta, Aptdo 02800, Chile.\\
$^2$ Departamento de F\'{\i}sica,  Facultad de Ciencias,\\
Universidad Central de Venezuela, A.P. 47270, Caracas 1041-A, Venezuela. \\
$^3$ Departamento de F\'{\i}sica, Universidad Sim\'on Bol\'{\i}var,\\
Apartado 89000, Caracas 1080-A, Venezuela}

\ead{$^a$gabriel.abellan@ciens.ucv.ve;
$^b$camilo.lasheras@uantof.cl; $^c$maria.garciadelmoral@uantof.cl; $^d$jpena@fisica.ciens.ucv.ve; $^e$alvaro.restuccia@uantof.cl}

\begin{abstract} Supermembrane compactified on a $M_9\times T^2$ target space is globally described by the inequivalent classes of torus bundles over torus. These torus bundles have  monodromy in $SL(2,Z)$ when they correspond to the nontrivial central charge sector and they are trivial otherwise. The first ones contain eight inequivalent classes of M2-brane bundles which  at low energies, are in correspondence with the eight type II gauged supergravities in $9D$. The relation among them is completely determined by the global action of T-duality which interchanges topological  invariants of the two tori. The M2-brane torus bundles are invariant under $SL(2,Z)\times SL(2,Z) \times Z_2$.  From the effective point of view, there is another dual invariant theory, called Double Field Theory which describe invariant actions under $O(D,D)$. Globally it is formulated in terms of doubled $2D$ torus fibrations over the spacetime with a monodromy given by $O(D,D,Z)$.  In this note we discuss T-duality global aspects considered in both theories and we emphasize certain similarities between both approaches which could give some hints towards a deeper relationship between them.
\end{abstract}

\section{Introduction}
 In the context of  M-theory, String Theory or Effective Field Theories the search for T-dual/U-dual invariant theories is a relevant goal. In this note we want to  discuss and briefly compare aspects of the bundle construction of two different duality invariant theories: Supermembrane theory, also called M2-brane theory and Double field theory (DFT). We consider of interest to understand whether or not there is any hint of a  connection  between them. In \cite{3hull99, 4hull04} it was argued that a fundamental formulation of string/M-theory should exist in which T- and U-duality symmetries are manifest from the start.  In particular, it was argued that many massive, gauged supergravities cannot be naturally embedded in string theory without such a framework \cite{stw, 6hullreid, reids, 7samtleben}.

Supermembranes are elements of M-theory, for that reason, dualities should appear as discrete symmetries of the theory. In this note we will restrict our analysis  to the case of a supermembrane theory formulated on $M_9\times T^2$. It contains two different topological sectors one with irreducible wrapping and a second one that is reducible. The theory corresponding to the irreducible wrapping sector called 'supermembrane with central charges' \cite{mor}. It is consistently defined at quantum level \cite{bgmr}, in distinction with the reducible case \cite{dwln, dwpp}. At low energies it is described by the type II gauged supergravity theories \cite{gmpr, ultimo}. The supermembrane without  the central charge condition corresponds to the type II maximal supergravity in nine dimensions. The mass operator and the hamiltonian are U-dual invariant \cite{gmpr} and when supermembrane is dimensionally reduced to string theory T-duality for supermembranes agree with the standard one in string theory compactified on a circle \cite{gmpr3}. Moreover, for the supermembrane with nontrivial central charge, it was  found  all the inequivalent classes of torus bundles with $SL(2,Z)$ monodromy that describes globally the theory, \cite{ultimo}. The M2-brane torus bundles are classified by their coinvariants of the base and the fiber and their associated monodromies. T-duality action determines the structure of the T-dual M2-brane bundle. M2-brane torus bundles with parabolic monodromy are the unique class of bundles invariant  not only locally but also globally, i.e. preserving the coinvariants. This implies that at low energy the T-dual of a type IIB parabolic gauged supergravity is another type IIA parabolic gauged supergravity. The precise T-duality  action on the classification of the M2-brane bundles: elliptic, parabolic, hyperbolic or with trombone monodromy is done in \cite{ultimo}. The construction is always globally well defined, hence geometric.

DFT is an effective field theory that describes sigma models on toroidal compactifications in which the number of spatial coordinates has been doubled to mimic the effect of winding modes for string theory. These coordinates form a doubled torus $T^{2d}$  with $d$-coordinates conjugate to the momenta and the other $d$-coordinates conjugate to the winding modes \cite{4hull04}. In order to preserve the physical number of propagating degrees of freedom of the theory a constrain is imposed. The global formulation of the theory is done in terms of torus fibrations over a spacetime manifold such that the transition functions are evaluated not only under diffeomorphims and gauge shifts but also under the T-duality group $O(d,d,{Z})\sim SL(d,Z)\times SL(d,Z)\times Z_2\times Z_2$ transformations \cite{lust}. The corresponding action is invariant under the duality transformations. String theory can be consistently defined in those backgrounds called non geometric backgrounds
\cite{4hull04, 3hull99}.  Such backgrounds can arise from compactifications with duality twists \cite{dh2} or from acting on geometric backgrounds with fluxes with T-duality \cite{3hull99, 4hull04, stw}. In special cases, the compactifications with duality twists are equivalent to asymmetric orbifolds which can give consistent string backgrounds \cite{12flournoyw, 13kawai, 18cederwall, dhgt, hullng}. Some global aspects of T-duality were  analyzed  in
\cite{24hull-tfolds, stw, dh2, 12flournoyw, 13kawai, 18cederwall, dhgt, lust} . These torus bundles have an $O(d,d,Z)$ monodromy of the doubled torus over the base spacetime manifold. Examples of generalized T-folds \cite{24hull-tfolds, 12flournoyw} can be obtained by constructing torus fibrations over base manifolds with singularities or with non-contractible cycles, like for example over a torus.\newline
The paper is structured in the following way: in section 2 we are going to review the supermembrane theory description in terms of bundles: we indicate the local (Hamiltonian and mass operator) and the T-duality global action  on the Supermembrane theory. In section 3 we review very basic facts of DFT in particular focusing on its global description to introduce next section. In section 4 we compare both approaches indicating differences and similarities and we present our conclusions.


\section{Supermembrane theory on a symplectic torus bundle}
Let us consider a supermembrane theory with central charges formulated in the Light Cone Gauge (LCG) on a target space $M_9\times T^2$ \cite{mor, gmmpr, sculpting}. $X^{n}$ are the embedding maps $\Sigma\to M_9$ where ${n}=3,\dots,9$ and $X=X^1+iX^2$  from $\Sigma\to T^2$. They are scalars parametrizing the transverse coordinates of the supermembrane in the target space. The moduli of the target space 2-torus $T^2$ are  the radius ${\mathtt{R}}$, and the complex Teichm\"uller parameter $\tau$.
The supermembrane has a nontrivial embedding over all the coordinates of the target. We define a matrix $\mathbb{W}=\begin{pmatrix} l_{1}& l_{2}\\
               m_{1} & m_{2}
 \end{pmatrix}$ whose entires are the winding numbers of the embedded supermembrane. When the wrapping of the supermembrane is irreducible $det \mathbb{W}=\mathtt{n}\ne 0$  the theory has discrete spectrum \cite{bgmr}. For the case $det \mathbb{W}=0$ the spectrum is continuous \cite{dwln}. The topological condition $\mathtt{n}\ne 0$ defines a principal line bundle with first chern class $c_1=\mathtt{n}$ that algebraically implies the existence of a non-vanishing central charge in the supersymmetric algebra of the supermembrane. This sector of the wrapped supermembrane theory we denote as supermembrane with central charges.
The Hamiltonian describing it is found in \cite{sculpting, gmpr, mor1}:
{\small\small
\begin{equation}
\begin{aligned}
H=&\int_{\Sigma}{T_{\mathtt{{M2}}}^{2/3}}\sqrt{\rho}\left[\frac{1}{2}(\frac{P_{n}}{\sqrt{\rho}})^2
+\frac{1}{2}\frac{P\overline{P}}{\sqrt{\rho}}+\frac{T_{\mathtt{{M2}}}^{2}}{4}\{X^{m},X^{n}\}^2+\frac{T_{\mathtt{{M2}}}^{2}}{2}(\mathcal{D}X^{n})(\mathcal{\overline{D}}X^{n})\right]+\\
+&\int_{\Sigma}{T_{\mathtt{{M2}}}^{2/3}}\sqrt{\rho}\left[+\frac{T_{\mathtt{{M2}}}^{2}}{4}(\mathcal{F}\overline{\mathcal{F}})\right]+({\mathtt{n}}^2 \, Area_{T^2}^2)+\\
+&\int_{\Sigma}{T_{\mathtt{{M2}}}^{2/3}}\sqrt{\rho}\left[-\overline{\Psi}\Gamma_{-}\Gamma_{n}\{X^{n},\Psi\}
-\frac{1}{2}\overline{\Psi}\Gamma_{-}\overline{\Gamma}\mathcal{D}\Psi\}-\frac{1}{2}\overline{\Psi}\Gamma_{-}\Gamma\overline{\mathcal{D}}\Psi\}\right]\\
+&\int_{\Sigma}\sqrt{\rho} \, {L} \left[\frac{1}{2}\overline{\mathcal{D}}(\frac{P}{\sqrt{\rho}})+\frac{1}{2}\mathcal{D}(\frac{\overline{P}}{\sqrt{\rho}})+\{X^{n},\frac{P_{n}}{\sqrt{\rho}}\}-\{\overline{\Psi}\Gamma_{-},\Psi\}\right],
\end{aligned}
\end{equation}}

\noindent where $L$ is a Lagrange multiplier, ${T_{\mathtt{{M2}}}}$ is the 11D tension of the supermembrane, $\rho$ is the determinant of the spatial part of the non-flat worldvolume two torus of supermembrane $\Sigma$. The symplectic bracket is defined as $\{\textsf{A},\textsf{B}\}=\omega^{ab}\partial_a \textsf{A}\partial_b \textsf{B}$ with $\omega^{ab}=\frac{i\epsilon^{ab}}{2\sqrt{\rho}}$ and  $a,b=z,\overline{z}$ the complex coordinates defined on $\Sigma$. The densities $P_{n}$ are the canonical momenta associated to the $X^{n}$ and $P$ those of $X$. $\Psi$ are scalars on the worldvolume but an $SO(7)$ spinor on the target space. $\Gamma_n$ are seven Gamma matrices and  $\Gamma= \Gamma_1+i\Gamma_2$, denoting by $\overline{\Gamma}$ its complex conjugate. $\mathcal{F}=D\overline{A}-\overline{D}A+\{A,\overline{A}\}.$  is a symplectic curvature defined on the base manifold with $A$ a connection under the infinitesimal symplectomorphism transformation $\delta_{\epsilon}A=\mathcal{D}\epsilon$. See \cite{mor1, gmmpr} for a detailed analysis. The symplectic covariant derivative is defined as $\mathcal{D}\bullet=D\bullet+\{A,\bullet\}$ with $D\bullet=e_r^a\partial_a\bullet$ a rotated covariant derivative defined in terms of a zwei-bein $e_r^a$ as, \cite{sculpting, gmpr} $e_r^a:=-2\pi {\mathtt{R}}(l_r+m_r\tau)\Theta_{sr}\omega^{ba}\partial_b\widehat{ X^s},$
with $r,s=1,2$ and being $d\widehat{X}$ are the harmonic one-form basis defined on $\Sigma$.
\paragraph{Symmetries of the Theory}
The Hamiltonian is invariant the residual symmetry under Area Preserving Diffeomorphisms (APD) connected and not connected to the identity. The theory is also invariant under two different $SL(2,Z)$ discrete symmetries: There is a $SL(2,Z)_{\Sigma}$ associated to the invariance under the change of the basis of the harmonic one forms defined on $\Sigma$ \cite{sl2z} and the windings,
\begin{equation}
d\widehat{X}\to Sd\widehat{X},\quad \mathbb{W}\to S^{-1}\mathbb{W} \,,
\end{equation}
with $S\in SL(2,Z)$. This dependence is encoded in the symplectic covariant derivative of the Hamiltonian through the matrix $\Theta\in SL(2,Z)$ \cite{sculpting}. There is a second $SL(2,Z)_{T^2}$ global symmetry, the S-duality transformation, associated to an invariance of the mass operator related to the target 2-torus $T^2$,
\begin{equation}
\label{transformations}
\tau \to \frac{a\tau +b}{c\tau +d}, \quad {\mathtt{R}} \to  {\mathtt{R}}|c \tau +d|, \quad A \to A e^{i\varphi_{\tau}} \, ,
 \mathbb{W} \to \begin{pmatrix} a & -b\\
                        -c & d
 \end{pmatrix} \mathbb{W}\,,\quad
{\mathbf{Q}} \to \Lambda{\mathbf{Q}},
\end{equation}
where $\Lambda=\left( \begin{array}{cc} a & b \\ c & d \end{array}\right) \in SL(2,{Z})$, $e^{-i\varphi_{\tau}}=\frac{c\tau+d}{|c\tau+d|}$ and ${\mathbf{Q}}=\left(\begin{array}{cc} p \\ q \end{array}\right)$ the Kaluza Klein (KK) charges of the supermembrane propagating on the target 2-torus $T^2$ considered.
\paragraph{Bundle description of the Supermembrane}
The embedding description did in \cite{mor, gmmpr, sculpting} can be also understood in terms of a symplectic torus bundle over a torus with monodromy in $SL(2,Z)$. This global formulation show topological invariants that carry physical information. The total bundle space $E$ is defined in terms of a fiber $F=M_9\times T^2$ and $\Sigma$ as the base manifold non trivially patched. The structure group $\mathtt{G}$ is the symplectomorphisms group  leaving invariant the canonical symplectic structure in $T^2$. The action of $\mathtt{G}$ on $F$ produces a $\pi_0(\mathtt{G})$-action on the homology and cohomology of $F$. In \cite{ultimo}, the authors give a geometrical interpretation of the KK charges in terms of  the $H_1(T^2)$ charges. The \textit{monodromy} of the bundle ${\mathcal{M}}$ is defined as
\begin{equation}
{{\mathcal{M}}}:\pi_1(\Sigma) \to \pi_0(\mathtt{G}),\quad \textrm{with} \quad \mathtt{G}=Symp(T^2) \quad \textrm{and} \quad \pi_0(\mathtt{G})=SL(2,Z)\, .
\end{equation}
Consequently ${{\mathcal{M}}}=\Lambda^{\gamma}\in SL(2,Z)$  and it acts on the homology basis of the $T^2$ target torus with  $\gamma=\gamma_1+\gamma_2$ an integer defined in terms of the integer basis of the fundamental group $\pi_1(\Sigma)$. The global symmetries of the theory become restricted by the monodromy.
The symplectic connection $A$ defined on the base manifold transforms with the monodromy as $dA\to dA e^{i\varphi_{{\mathcal{M}}}}$ where ${\varphi}_{{{\mathcal{M}}}}=\frac{c\tau+d}{\vert c\tau+d\vert}$ is a discrete monodromy phase for a given modulus $\tau$. In distinction the Hamiltonian is invariant.
The torus bundles with a given monodromy ${\mathcal{M}}$ are classified according to the elements of the twisted second cohomology group $H^2 (\Sigma, Z_{{\mathcal{M}}}^2)$ of the base manifold $\Sigma$ with coefficients on the module generated by the monodromy representation acting on the homology of the target torus \cite{khan}. There is a bijective relation between the elements of the coinvariant group $C_F=\{C_{\mathbf{a}}\}, {\mathtt{a}}=1,\dots,j$ associated with a particular monodromy group ${\mathcal{M}}_G$. A coinvariant class in the KK sector is given by
\begin{equation}
C_{F}=\{{\mathbf{Q}}+({\mathcal{M}}_g-\mathbb{I})\widehat{{\mathbf{Q}}}\} \, ,
\label{coinvariantclasskk}
\end{equation}
for any element $\mathcal{M}_g\in {\mathcal{M}}$, and $\widehat{{\mathbf{Q}}}$ is any arbitrary element of the KK sector. There is also an induced action on the cohomology of the base manifold, which corresponds to a '\textit{monodromy}' group of the winding sector  ${\mathcal{M}}_G^*$.
A coinvariant class in the winding sector is given by
\begin{equation}
C_B=\{{\mathbf{W}}+({\mathcal{M}}_g^*-\mathbb{I})\widehat{{\mathbf{W}}}\},
\end{equation}
with $\mathcal{M}_g^*\in {\mathcal{M}}^*=\Omega {\mathcal{M}} \Omega^{-1}$ with $\Omega= \begin{pmatrix}-1& 0\\ 0 & 1 \end{pmatrix}$ , and ${\mathbf{W}}=\left(\begin{array}{cc} l \\ m \end{array}\right)\in H^1(\Sigma)$. The ${\mathcal{M}}^*$ acts on the fields which define the Hamiltonian, through the matrix $\Theta=(V^{-1}{\mathcal{M}}^*V)^T$  that appears in the symplectic covariant derivatives $\mathcal{D}_r$ \cite{sculpting}. ${\mathcal{M}}$ and ${\mathcal{M}}^{*}$ groups lie in the same equivalence class but their respective coinvariants classifying bundles are not equivalent.
%
%
Then in order to specify the physical content of the M2-brane on a symplectic torus bundles one also needs to determine $(C_F, C_B, \mathcal{M}_G)$. Each coinvariant class is invariant under the action of any element of the monodromy group $\mathcal{M}_g\in {\mathcal{M}}$. Among its elements we have the orbits of any element of the class. So the coinvariant class may be considered itself as a class of orbits under the action of ${\mathcal{M}}$. Given a symplectic torus bundle the Hamiltonian of the theory is defined for any orbit of the coinvariant class.
%
%
%
\subsection{T-duality for supermembrane theory torus bundles}
The T-duality transformation acts on the Hamiltonian $H$ and the mass operator ${\bf M}^2$ and it has also a action on the structure of the bundle consequently on their topological  invariants describing the M2-brane theory. Globally the T-duality transforms a bundle into a dual one, by interchanging the cohomological charges of the torus base manifold into the homological charges defined on the torus fiber with dual moduli. T-duality also interchanges the coinvariant class of the base and the fiber in the dual T-bundle, $(C_F,C_B)=(\widetilde{C}_B, \widetilde{C}_F),
$ where we denote by tildes the quantities in the dual bundle. In general this transformation becomes non linear. At low energies this fact will be reflected in the change of the gauging group associated to the corresponding dual supergravity. In order to analyze it with more detail, we will consider separately the cases of trivial and non trivial monodromy:
\begin{itemize}
\item {\em Trivial monodromy}: In this case ${\mathcal{M}}=\mathbb{I}$, and the coinvariant classes, which classify the inequivalent torus bundles, have only one element $\mathbf{Q}$ in the KK sector and one element $\mathbf{W}$ in the winding sector. The duality transformation on the symplectic torus bundle has an action on the charges but also on the geometrical moduli. Following the notation of \cite{ultimo}, we define dimensionless variables $\mathcal{Z}=(T_{\mathtt{M2}} \mathtt{A} Y)^{1/3}$ where $\mathtt{A} = (2\pi \mathtt{R})^2 Im\tau$ is the area of the target torus and $Y=\frac{\mathtt{R} Im\tau}{\vert q\tau -p\vert}$ is a variable proportional to the $\mathtt{R}$ radius of the complex torus. The T-duality transformation is given by:
\begin{equation}
\label{tt1moduli}
\textrm{The moduli}: \quad \mathcal{Z}\widetilde{\mathcal{Z}}=1,\quad\widetilde{\tau}=\frac{\alpha \tau+\beta}{\gamma\tau +\alpha}.
\end{equation}
\begin{equation}
\label{tt1charges}
\textrm{The charges}: \quad \widetilde{\mathbf{Q}}=\mathcal{T} \mathbf{Q},\quad \widetilde{\mathbf{W}}=\mathcal{T}^{-1}\mathbf{W},
\end{equation}
\noindent with $\mathcal{T}=\left( \begin{array}{cc} \alpha & \beta \\ \gamma & \alpha \end{array}\right) \in SL(2,{Z})$. The symplectic torus bundles are classified in this case by two integers, the elements $Z\otimes Z$ and they are in one to one correspondence with the $U(1)\times U(1)$ principle bundle over the base manifold. Since the monodromy is trivial, the structure group may reduce to the group of symplectomorphism homotopic to the identity. The dual transformation is then completed by the transformation of the moduli as given in (\ref{tt1moduli}).\newline

\item {\em Non trivial monodromy:} In this case, the monodromy group $\mathcal{M}$ is non trivial and it is an abelian subgroup of $SL(2,Z)$.  The T-dual transformation maps as before coinvariant classes on the KK sector onto coinvariant class in the winding sector and each class denoted as $[\quad ]$ contains several elements.
Hence the map between the coinvariant classes is only determined by $\mathcal{T}$ which is constructed from one element of each class $\mathbf{Q}$ and $\mathbf{W}$ respectively. The monodromy ${\mathcal{M}}$ can be parabolic, elliptic or hyperbolic, which can be linearly or non-linearly realized (trombone symmetries). The T-duality transformation is given by (\ref{tt1moduli}) and for the charges:
\begin{equation}
\label{tt2charges}
[\widetilde{\mathbf{\mathbf{W}}}]=\mathcal{T} [\mathbf{Q}],\quad [\widetilde{\mathbf{Q}}]=\mathcal{T}^{-1}[\mathbf{W}],\quad
{\mathcal{M}}\stackrel{\Omega}{\rightarrow} \mathcal{M}^*,\quad
(C_F,C_B) \to (\widetilde{C}_B,\widetilde{C}_F).
\end{equation}
The variables associated to the geometric moduli $\mathcal{Z},Y$ and their duals are invariant on an orbit generated by ${\mathcal{M}}$ contained in the respective coinvariant class, provided that $\tau$ and $\mathbf{Q}$ transform as in (\ref{transformations}). The symmetry of the Hamiltonian related to the basis of harmonic one-forms of $\Sigma$ \cite{sl2z} allows to define the class of orbits associated to the winding matrices $[\mathbf{W}]$. T-duality defines a nonlinear transformation on the charges of the supermembrane since $\mathcal{T}$ is constructed from them, in distinction with the $SL(2,Z)_T$ action on the moduli which is a linear one. The condition $\mathcal{Z}\widetilde{\mathcal{Z}}=1$ ensure that $(\textrm{T-duality})^2= {I}$. This transformation becomes a symmetry for $\mathcal{Z}=\mathcal{\widetilde{Z}}=1$ which imposes a relation between the tension, the moduli and the KK charges of the wrapped supermembrane, $
 T^0_{\mathtt{M2}}=\frac{\vert q\tau-p\vert}{{\mathtt{R}}^3 (Im\tau)^2}
$.  Given the values of the moduli and the charges, the allowed tension is fixed $T^0_{\mathtt{M2}}$. For $\mathcal{Z}=1$ the Hamiltonian and the mass operator of the supermembrane with central charges are invariant under T-duality:
\begin{equation}
{\bf  M}^2 = (T^0_{\mathtt{M2}})^2 {\mathtt{n}}^2 \mathtt{A}^2 + \frac{k^2}{Y^2}+ (T^0_{\mathtt{M2}})^{2/3}H =\frac{{\mathtt{n}}^2}{\widetilde{Y}^2}+ (T^0_{\mathtt{M2}})^2 k^2 \widetilde{\mathtt{A}}^2+ (T^0_{\mathtt{M2}})^{2/3}\widetilde{H},
\end{equation}
with $H=\widetilde{H}$. The mass operator of the supermembrane is U-dual invariant.
Thought the mass operator is always invariant, under T-duality only the parabolic M2-brane bundles class is invariant since the action of T-duality is linear and the structure of the coinvariants is preserved, so the bundle. For the rest of the monodromies contained in $SL(2,Z)$ with linear or non linear realization the action of T-duality is non linear, the monodromy dual is always in the same monodromy class but the bundle is classified by different coinvariants \cite{ultimo}.
\end{itemize}
%
\subsection{Supermembrane theory and gauged supergravity in 9D} Scherk-Schwarz compactifications of supergravity may be expressed in terms of principal fiber bundles over circles with a twisting given by the monodromy \cite{hull-massive, pope}. The background possesses  a group of global isometries $G$ associated to the compactification manifold over which it is fibered. In the type II gauged supergravities in 9D, the monodromies are associated to the $GL(2,{R})= SL(2,{R})\times {R}^{+}$ global symmetry group. In the $SL(2,{R})$ sector, there are three  inequivalent classes of theories, corresponding to the hyperbolic, elliptic and parabolic $SL(2,{R})$ conjugacy classes, see \cite{hull-massive}. The gauging of the $R^+$ scaling symmetry called trombone gives supergravities without lagrangians \cite{samtleben-trombone, riccioni}. At quantum level this last symmetry corresponds to an $SL(2,Z)$ non linearly realized and gives rise to a different symplectic torus bundle in comparison to the previous constructions in terms of linear representations \cite{gmpr3, ultimo}. Indeed, only when the monodromy is parabolic, the T-duality action is linear meanwhile for the elliptic and hyperbolic case, T-duality does not commute with the monodromy but forms a unique class of torus bundles in the type IIA side with an scaling symmetry $A(1)$. The T-duality  of M2-brane trombone bundles maps also in the 'type IIA side' into two inequivalent classes of M2-brane dual trombone  bundles. These classes are in perfect agreement with the eight type II gauged supergravities, four in the type IIB (elliptic, parabolic, hyperbolic and trombone) and other four in the type IIA (two trombone, one parabolic and one non abelian with group $A(1)$ \cite{bergshoeff, etjjto}.
\section{Global Aspects of T-duality in Double Field Theory}
We review here very general facts of the global description of DFT. It is a reformulation of supergravity constructed in such a way that it  is T-dual invariant and it makes explicit the duality group $O(D,D)$ of the theory \cite{DFT2}. So far, it has been studied extensively for the bosonic sector of the theory. T-duality is a transformation that changes moduli at the same time that KK charges are interchanged with the winding modes. Winding modes are a property of extended objects, hence to introduce T-duality invariance into an effective field theory, the spatial coordinates are duplicated $(x_D,\widetilde{x}^D)$  so that $\widetilde{x}^i, i=1\dots D$ are conjugated to the winding modes $\omega^i, i=1\dots D$ \cite{tseytlin} and for consistency of the theory many algebraic structures become enlarged, see for a review, \cite{DFT}. The worldsheet description of DFT on tori backgrounds has been studied in \cite{lust, grana}.
The string action on a toroidal general background is the following \cite{kugo},
	\begin{eqnarray}
	S=-\frac{1}{4\pi}\int_0^{2\pi}d\sigma \int d\tau \Big( \sqrt{\gamma}\gamma^{\alpha\beta}\partial_\alpha X^i\partial_\beta X^j G_{ij}+ \epsilon^{\alpha\beta}\partial_\alpha X^i\partial_\beta X^j B_{ij}\Big), \label{4.1}
	\end{eqnarray}
where $\gamma_{\alpha\beta}$ is the induced worldsheet metric, $B_{ij}$ is a constant target space 2-form and $G_{ij}$ is the background metric. The associated Hamiltonian can be written in terms of a $2D\times 2D$ generalized metric ${\mathcal{H}}_G({E})$ constructed in terms of the background metric  $G_{ij}$ and the two-form $B_2$ in a non-trivial way \cite{DFT2}
	\begin{eqnarray} \label{4.17}
	{\mathcal{H}}_{G}({E})=\left[ {\begin{array}{cc} G_{ij}-B_{ik}G^{kl}B_{lj} & B_{ik}G^{kj} \\ -G^{ik}B_{kj} & G^{ij} \end{array}} \right]\,,
	\end{eqnarray}
such that worldsheet hamiltonian density ${H}$ can be expressed as
	\begin{eqnarray}\label{4.18}
	{H}=\frac{1}{2}Z^T {\mathcal{H}}_{G}({E})Z+N+\bar{N} \,,
	\end{eqnarray}
where $\displaystyle N$ y $\displaystyle \bar{N}$ are the left and right number operators and $\displaystyle Z=\left( {\begin{array}{c} \omega^i \\ p_i\end{array}} \right) $ is a generalized momentum that includes the KK momentum $p_i$ and the winding modes $\omega^i$, and a background matrix ${E}_{ij}$ as
	\begin{eqnarray}
	{E}_{ij}\equiv G_{ij}+B_{ij}=\left[ {\begin{array}{cc} {E}_{mn}& 0 \\ 0 & g_{\mu\nu} \end{array}} \right], \label{4.5}
	\end{eqnarray}
with ${E}_{mn}=G_{mn}+B_{mn}$, $G_{mn}$ is the flat metric for the torus, $B_{mn}$ the Kalb-Rammond 2-form components over the $n$-torus and ${g_{\mu\nu}}$ is background metric on a target space $M^{D-n-1,1}$.  The hamiltonian density ($\ref{4.18}$) is $O(D,D)$ invariant when it is induced a transformation law on the generalized metric
$
	{\mathcal{H}}_G({E}')=h\,{\mathcal{H}}_G({E})\,h^T \,, \label{4.28}
$
where $h$ is an element of $O(D,D,{R})$ group and verifies $h\eta h^T=\eta$, with $\eta=\begin{pmatrix} 0 & \mathbb{I}_{D\times D} \\ \mathbb{I}_{D\times D} & 0 \end{pmatrix}$,  the $O(D,D,{R})$ invariant metric.
%
Due to the charge quantization condition, $\omega^m$ and $p_m$ are restricted to take discrete values. This implies that the symmetry group is restricted to be its arithmetic subgroup $O(n,n,{Z})$, the T-duality group of string theory. An element of the T-duality group $h\in O(n,n,{Z})$ can be expressed in terms of an $O(D,D)$ representation for that reason one can consider $O(D,D)$, as the global T-duality group.

The idea behind DFT is to find a T-duality invariant action effective field theory describing supergravity. To this end authors in \cite{giveon, Schwarz}, construct an $O(D,D)$ tensor $\mathcal{H}_{MN}$, where $M,N=1,\dots,2D$ are $O(D,D)$ curved indices, of the target space metric $g_{ij}$, the Kalb-Rammond 2-form $b_{ij}$ and the dilaton $\phi$, combined as in (\ref{4.17}). The dilaton  can be expressed as an $O(D,D)$ singlet
$
	e^{-2d}=\sqrt{g}e^{-2\phi} . \label{4.44}
$
 A new set of coordinates $\tilde{x}_i$ conjugated to the winding modes $\omega^i $ are defined and with them some new generalized coordinates $ X ^ M=(\tilde{x}^i,x_i) $ considering the original and the dual ones are introduced. These coordinates induce the generalized derivatives $\tilde{\partial_M}=\Big( \frac{\partial}{\partial\tilde{x}_i},\frac{\partial}{\partial x^i}\Big)$. Additionally the fields depends on this generalized coordinates $\mathcal{H}_{MN}(X),d(X)$ and it is possible to construct the $O(D,D)$ invariant action as a generalization of Einstein Hilbert terms \cite{Hohm-Hull-Zwibach},
	\begin{eqnarray}
	S &=& \int d^{2D}Xe^{-2d}\mathcal{R}, \label{4.51}
	\end{eqnarray}
where the generalized curvature is given by
	\begin{eqnarray}	
	\mathcal{R} &=& 4\mathcal{H}^{MN}\partial_M\partial_Nd - \partial_M\partial_N\mathcal{H}^{MN} - 4\mathcal{H}^{MN}\partial_Md\partial_Nd + 4\partial_M\mathcal{H}^{MN}\delta_Nd, \nonumber\\
	&+& \frac{1}{8}\mathcal{H}^{MN}\partial_M\mathcal{H}^{KL}\partial_N\mathcal{H}_{KL} - \frac{1}{2}\mathcal{H}^{MN}\partial_M\mathcal{H}^{KL}\partial_K\mathcal{H}_{NL}. \label{4.52}
	\end{eqnarray}	
 In the same way one generalizes the rest of supergravity terms and to do it is also necessary to modify several mathematical structures as the Lie derivative among others. Since the physical degrees of freedom have been doubled, a constraint is imposed to preserve the correct number of degrees of freedom. The strong constraint closes a generalized Lie derivative gauge algebra
	\begin{eqnarray}
	\eta^{MN}\partial_M(A)\partial_N(B)=\partial^M(A)\partial_M(B)= \partial^M\partial_M(AB)=0,  \label{4.54}
	\end{eqnarray}
where $\eta_{MN}$  is the $O(D,D)$ invariant metric, $A$ and $B$ represents any DFT field or generalized Lie Derivative Parameter $\xi^M = (\tilde{\lambda}_i,\lambda^i$)  defined by
	\begin{eqnarray}
	\mathcal{L}_{\xi}A_M &\equiv& \xi^P\partial_PA_M + (\partial_M\xi^P - \partial^P\xi_M)A_P, \label{4.56} \\
	\mathcal{L}_{\xi}B^M &\equiv& \xi^P\partial_PA^M + (\partial^M\xi_P - \partial_P\xi^M)B^P. \label{4.57}
	\end{eqnarray}
	
	After imposing the constraint, the field configuration depends only of the coordinates of the $D$-dimensional subspace, which in general is a linear combination of the original $x^i$ and the dual coordinates $\tilde{x}_i$.

\paragraph{Global description} String theory can be consistently defined in backgrounds in which the transition functions between the patches considers not only diffeomorphisms and gauge transformations but also T-duality transformations \cite{4hull04}. Those backgrounds that include T-duality transformations are called non geometrical. If one restricts to the case in which  each transition function patches the charts with an element of $GL(n,Z)=SL(n,Z)\times Z_2 \subset O(n,n,Z)$, the large diffeomorphisms group acting on the fiber, then the background is geometrical. However, this is not the most general case since in string theory T-duality maps the transitions functions $S\in GL (n,{Z}) $ into $S'= gSg^{-1}$ with $g\in O(n,n,{Z})$ and in general $S'\in O(n,n,{Z})$ so the background becomes non-geometric.  M-theory, as a theory of unification, contains the dualities as symmetries of the theory and DFT aims to be an effective description of M-theory, then it is natural to consider bundles whose  transition functions are defined in $O(D,D,{Z})$. In the DFT global formulation that we are considering, the target space $M $ is locally a torus bundle with fiber $T^D$ over the base manifold $N$, is extended by considering another $T^D$ on the fiber, where both are subspaces of a doubled torus $T^{2D}$ containing the original one and its dual, forming an extended target space $\Tilde{M}$ that can be interpreted as a torus bundle with fiber $T^{2D}$ over the base manifold $ N $. Each of the two torus has an $SL(D,Z)$ invariance associated to its mapping class group. $\tilde{M}$ is a $2D$ torus bundle with a well defined monodromy in $O (D,D,{Z})\subset GL(2D,{Z})$, since it is contained on the large diffeomorphims group of the $T^{2n}$. If $M$ is a geometric background and we have an $O(D,D,{Z})$ monodromy on the fiber, then we have a fibration of $\tilde{M}$ over $M$ \cite{4hull04}. However in general only  locally one can define the $T^D$ fibration with a $G$ and $B$ defined in the internal torus consequently  there is no well defined global geometry on the physical space-time: it is locally the product of a $D$-torus embedded in $T^{2D}$ and a sector of the base manifold $N$. This kind of construction which comes from torus fibrations over a torus base where the torus fiber undergoes a monodromy lying in the perturbative duality group $O(D,D,{Z})$, were called monodrofolds \cite{flournoy} or T-folds \cite{12flournoyw,24hull-tfolds}.

Since the theory contains more degrees of freedom than those physical by choosing a polarization one selects the physical subspace forming a torus contained in the doubled $T^{2D}$ for each point of the base manifold $N$. T-duality acts on the bundle by changing the polarization, that is, by changing the physical subspace $T^D\subset T^{2D}$ and for each chart $U$ of the manifold $N$, there is a local space-time chart $U\times T^D$ embedded in the doubled space $U\times T^{2D}$, while for local T-folds these do not form a spacetime manifold even though the entire doubled space forms a manifold which is a $T^{2D}$ bundle over the base manifold $N$.

\section{Discussion: A comparison of the global T-duality action in both approaches}
%

In this section we would like to compare the bundle construction of Supermembrane theory with respect to the one associated to the Double Field Theory. For the DFT case the bundle construction has been mainly explored in the T-dual invariant string theory framework. It contains nongeometric vacua solutions \cite{hullng, flournoy, 12flournoyw}. Both theories are clearly different, in spite of this, their bundle constructions have certain resemblances and differences that we would like to emphasize. To start with, DFT is an effective field theory while Supermembrane theory is part of M-theory. Supermembrane theory contains a topological sector (the one associated to the non trivial central charge) that can be consistently quantized.

The Supermembrane bundle that we analyze in this note has the M2-brane worldvolume torus as its base and the fiber is the compactified target space $M_9\times T^2$. The transition functions between the charts are given by the symplectomorphisms group preserving the symplectic 2-form on the base, and the bundle can be a trivial principal torus over a torus when the wrapped M2-brane has a vanishing central charge and a nontrivial symplectic torus bundles over a torus with monodromy in $SL(2,{Z})$ linearly or non linearly realized when the wrapping is irreducible,i.e. has a nontrivial central charge. The M2-brane torus bundle is specified by their coinvariants: $(C_B,C_F)$ which are defined in terms of two different monodromies $({\mathcal{M}}_B, {\mathcal{M}}_F)$  \cite{ultimo}. The hamiltonian is consistently defined on these M2-brane bundles. The global action of T-duality maps supermembrane class of bundles into dual ones, the homological charges of the torus of the fiber associated to the charges by the cohomological charges associated to the windings of the torus of the base. The tori geometry is not interchanged but the dual moduli of the fiber torus is obtained through the T-duality rules. The mass operator and hamiltonian are U-dual invariant.  The monodromy of the bundle is interchanged with the monodromy of the base \cite{ultimo}. The monodromy dual remains in the same conjugate class as the original one but in general the coinvariants that define the inequivalent classes of bundles, as we have discussed, are not preserved. The patching is always geometrical and consequently  the global picture well defined.  There is a $SL(2,{Z})_{\Sigma}\times SL(2,{Z})_{T^2}\times Z_2$ invariance on the M2-brane bundle, where the $Z_2$ action is associated to nontrivial discrete interchange of  global structures and topological invariants that we have already signaled.

On the other hand, in DFT global formulation, the fields are defined over the spacetime coordinates in distinction with the previous case. Its global description has been developed mainly inspired in the context of the corresponding duality invariant string worldsheet bundle.  The bundle contains the spacetime manifold as a base and a doubled  torus bundle as a fiber $T^{2D}$.  The transition functions in general are not restricted to be just the diffeomorphisms and the gauge invariance but the charts  are patched through elements of the T-duality group $O(D,D,{Z})$. In the case in which duality group is included the torus bundle has a non trivial monodromy. The fibration of the dual torus over the original one is only possible when the monodromy of the fiber belongs to the duality group $O(D,D,{Z})$.  In order to make contact between the two approaches, let us particularize to the case in which the DFT describes a $T^4=T^2\times T^2$ bundle with $O(2,2,{Z})$ monodromy.  The group of duality invariance of the theory $O(2,2,{Z})$ can be decomposed  as $SL(2,Z)_{\tau}\times SL(2,Z)_{\rho}\times Z_2\times Z_2$ where $SL(2,Z)_{\tau}$ is  associated $\tau$   the Teichm\"uller parameter of  the $T^2$ fiber, $SL(2,{Z})_{\rho}$ with  $\rho$ the complexified K\"ahler parameter defined in terms of the constant element of the 2-form $b$ and the area $A$ , that is $\rho=b+iA$ parameterizing the dual $\tilde{T^2}$ and the $Z_2$ discrete symmetries acts on both parameters as  follows \cite{lust},
\begin{equation}
\tau\Leftrightarrow \rho,\quad (\tau,\rho)\to (-\overline{\tau},-\overline{\rho}).
\end{equation}

In DFT theories, the T-duality action can be understood as a different choice of polarization which selects the physical subspace $T^D$ inside the doubled tori $T^{2D}$. It mixes the spatial coordinates $(X)$ with the doubled coordinates $(\widetilde{X})$. In those cases in which $T^{2D}= T^D\times T^D$ and the polarization selects as the physical tori the one associated with the spatial coordinates, T-duality intertwine the two tori, interchanging the geometry and not only the moduli under T-duality.  Consequence of this the two monodromies defined one associated to the $T^2_{\tau}$ and another $T^2_{\rho}$ over $N$ (on top of the $O(2,2,{Z})$ monodromy associated to the $T^4$ bundle  over the base manifold). These two monodromies become also interchanged under T-duality.
It seems clear that are resemblances between  both theories: Both theories realize similar symmetries in a different way, the worldvolume torus in the supermembrane and the dual 2-torus in the DFT are related to the winding charges. In both cases there are two tori on the bundle construction one of which is related to the torus target and the other to the winding modes, under T-duality in DFT these tori are interchanged but in the supermembrane case only the charges are interchanged but not the geometry. In both cases the two monodromies associated to each of the tori under T-duality are interchanged.  Both theories share the same discrete invariance $O(2,2,{Z})$ in the example considered but  differently realized since in the case of the Hamiltonian of the supermembrane, it involves the moduli in a nontrivial way. Finally, DFT is associated generically to non geometrical backgrounds since globally it is not well-defined, while the supermembrane formulation  -as it also happens for string theory- is always geometrical and globally well defined.

Since DFT has its roots in string theory and the five string theories have its origin in the supermembrane theory, a sector of M-theory, it is quite natural  to expect -even when there are differences-, that DFT describing the factorized doubled torus with a its monodromy in $ O(2,2,{Z})$ could be related to an effective description of the M2-brane torus bundle with monodromy in $SL(2,{Z})$. Obviously this hint will need to be confirmed by a more exhaustive study which is outside of the scope of the present note.

\section{Acknowledgements}  MPGM is supported by Mecesup ANT1656, Universidad de Antofagasta, (Chile). A.R.  and MPGM are partially supported by Projects Fondecyt 1161192 (Chile). J.M. Pe\~na is grateful to Universidad de Antofagasta for its hospitality and its partial support during part of the realization of this work.
%

\end{document}